\documentstyle[11pt,moriond,epsfig]{article}

\newcommand{\zpc}[3]{Z.\ Phys.\ {\bf C#1} (19#2) #3}

\newcommand{\npb}[3]{Nucl.\ Phys.\ {\bf B#1} (19#2) #3}
\newcommand{\prd}[3]{Phys.\ Rev.\ {\bf D#1} (19#2) #3}

\newcommand{\epjc}[3]{Euro.\ Phys.\ J.\ {\bf C#1} (19#2) #3}
\newcommand{\cpc}[3]{Comput.\ Phys.\ Commun.\ {\bf #1} (19#2) #3}

\def\be{\begin{equation}}
\def\ee{\end{equation}}
\def\bea{\begin{eqnarray}}
\def\eea{\end{eqnarray}}

\begin{document}
\begin{flushright}
 UCL/HEP 99-04\\
 May 1999
\end{flushright}
\vspace*{4cm}
\title{Heavy Flavours in Photoproduction at HERA}

\author{ Mark Hayes \\
(on behalf of the H1 and ZEUS collaborations)}

\address{Department of Physics and Astronomy,
University College London,\\
Gower Street,
LONDON,
WC1E 6BT }

\maketitle

\abstracts{
Differential cross sections, $d\sigma/d x_\gamma^{\rm OBS}$, for 
dijet photoproduction events with a tagged $D^*$ meson are presented,
 where $x_\gamma^{\rm OBS}$ is the fraction of the photon energy
contributing to the two highest transverse energy jets.
Results on open beauty cross sections 
(from semileptonic decays) compared to LO Monte Carlo predictions
are also presented. 
}

\section{Photoproduction}
The photon structure is being probed in new kinematic regimes at
HERA. Tagging of charmed mesons or semileptonic decays in
photoproduction events is providing a wealth of information on charm
in the photon and the processes involved in its production. Since 
photoproduction of charm and beauty is
much less affected by non-perturbative and higher-order effects than
hadroproduction cross sections, the \mbox{HERA} measurements can
provide important tests of the heavy-flavour production dynamics.

In photoproduction at leading order (LO), two types of process contribute to the
hard scatter: direct photoproduction, where the photon enters directly into
the hard subprocess, and resolved photoproduction, where the photon acts as a
source of partons, which in turn interact in the hard subprocess. At 
next-to-leading order (NLO), only the sum of direct and resolved processes is
unambiguously defined.

\section{Photoproduction of Charm}

\begin{figure}[tb]
  \epsfysize=10cm
  \epsfxsize=8cm
  \centerline{\epsffile{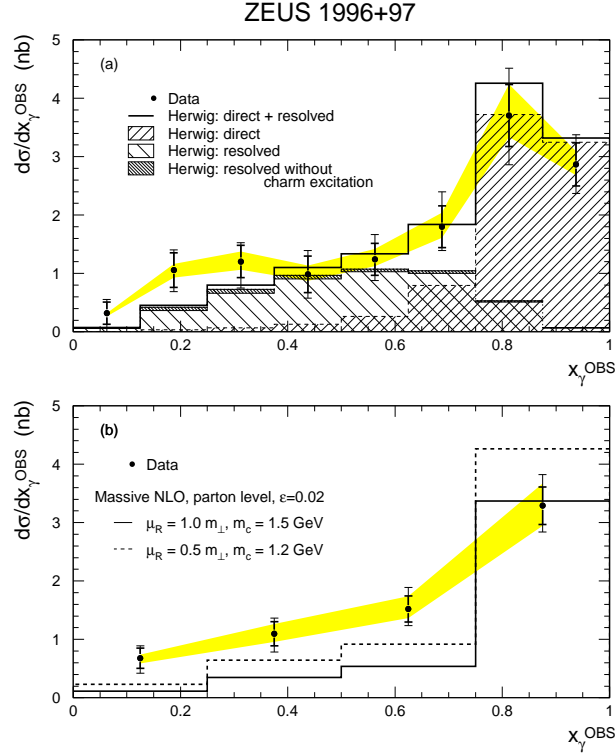}}
\caption[dummy]{\small \label{xgamma_dstar}
The $x_\gamma^{\rm OBS}$ distribution for tagged $D^*$ decays from
charm, beauty and light quarks. The points are ZEUS preliminary data
for the 1996 and 1997 data sample. The inner error bars are the
statistical errors and the outer error bars are the statistical and
systematic errors added in quadrature. The shaded curve represents the
uncertainty due to the energy scale of the jets. In (a) the shaded
histograms show the direct and resolved contributions as predicted by
HERWIG 5.9. The solid line is the two contributions added together.
In (b) the same data are shown (rebinned into four bins)
compared to a NLO calculation \cite{FMNR} with two choices of parameters.}
\end{figure}

H1 and ZEUS have published measurements of $D^*$ mesons in photoproduction
\cite{zeus98,h198}. In photoproduction events with two or more jets it 
is possible to define an observable 
\begin{equation}
x_\gamma^{\rm OBS} = \frac{\sum_{\rm jets}
E_T^{\rm jet}e^{-\eta^{\rm jet}}}{2y E_e},
\end{equation}
where the sum is over the two jets with highest transverse energy. In
LO, $x_\gamma^{\rm OBS}$ can be interpreted as an observable related
to the fraction of the photon's momentum participating in the
hard-scale interaction. At higher $x_\gamma^{\rm OBS} > 0.75$,  
the direct processes dominate, while at lower $x_\gamma^{\rm
OBS}$ the resolved processes dominate. 

ZEUS have published a measurement of $x_\gamma^{\rm OBS}$
with a tagged $D^*$ meson, for the kinematic regime, $Q^2 < 1~{\rm GeV}^2$, 
$130 < W_{\gamma P} < 280$, two jets with $E_T > 7,6$ GeV and $|\eta | < 2.4$, 
and a $D^*$ in the final state with $p_T > 3.0$ GeV and $|\eta | < 1.5$. 
This is shown in Figure~\ref{xgamma_dstar}. This shows a clear tail to low 
$x_\gamma^{\rm OBS}$, which in terms of
LO Monte Carlo programs can only be explained by a large (40\%)
component of LO-resolved processes. This is heavily dependent on the
amount of charm in the parton density function used. A NLO
calculation, performed in a factorization scheme where heavy quarks
are exclusively generated in the hard subprocess, cannot describe this 
long tail and falls below both the data and the LO Monte Carlo. 

\section{Photoproduction of Beauty}

H1 have released a preliminary beauty cross section and comparison with
the LO theory. They have used the $p_T^{\rm rel}$ method to
separate the beauty component. In this method, $p_T^{\rm rel}$ is the 
momentum of the muon transverse to the thrust axis of the constituent
particles (minus the muon) of the closest jet. 
The muons from beauty decays are expected to dominate at high 
$p_T^{\rm rel}$ because of the large beauty mass. The relative composition
of the data sample is determined from an unconstrained fit and amounts to 
$f_b = 51.4\pm 4.4\%$ (beauty), $f_c = 23.5 \pm 4.3 \%$ (charm) and 
$f_{\rm fake} = 23.5 \%$ (background, fixed).
For the kinematic
range, $Q^2 < 1~{\rm GeV}^2$, $0.1<y<0.8$, $p_T > 2.0$~GeV and
$35^\circ < \theta^\mu < 130^\circ$, H1 quote a preliminary visible cross
section of $\sigma^{\rm vis}_{ep \rightarrow e + b\overline{b} + X} =
0.93\pm 0.08 {}^{+0.21}_{-0.12}$~nb and also quote a predicted cross
section from AROMA~2.2 \cite{aroma} of $\sigma^{\rm vis}_{ep
\rightarrow e + b\overline{b} + X} = 0.191$~nb.  Thus the
preliminary measurement of the cross section of open beauty is up to
a factor of 5 higher than the LO theory as predicted by the Monte
Carlo AROMA. It should be noted that AROMA only produces
a direct component. Another additional resolved component should therefore be
taken into account in the comparison.

\begin{figure}[tb]
  \epsfysize=9cm
  \epsfxsize=10cm
  \centerline{\epsffile{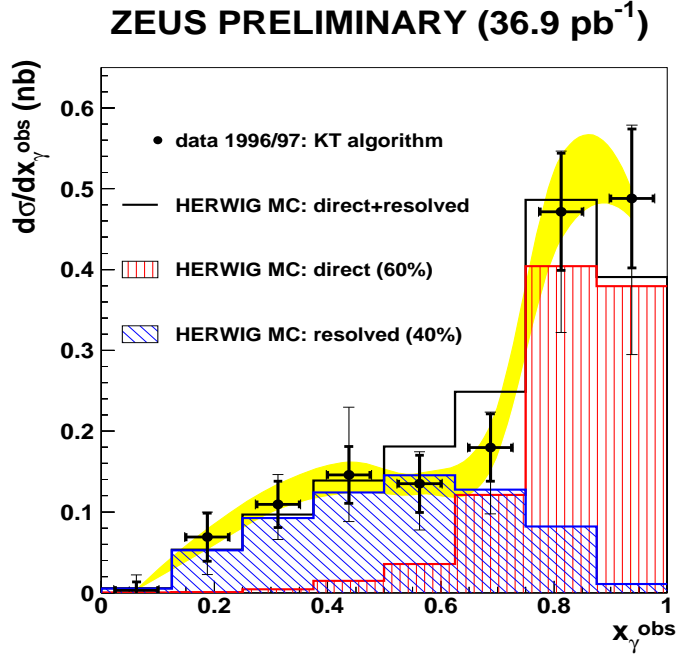}}
\caption[dummy]{\small \label{xg-sl}
The $x_\gamma^{\rm OBS}$ distribution for tagged semileptonic decays
as meaured in ZEUS. The points are ZEUS preliminary
1996 and 1997 data. The inner error bars are the statistical
errors and the outer error bars are the statistical and systematic
errors added in quadrature. The shaded curve represents the
uncertainty due to the energy scale of the jets. The shaded histograms 
show the direct and
resolved contributions for charm, beauty and light quarks 
as predicted by HERWIG 5.9. The solid line shown is the
sum of the contributions.}
\end{figure}

\begin{figure}[tb]
  \epsfysize=9cm
  \epsfxsize=10cm
  \centerline{\epsffile{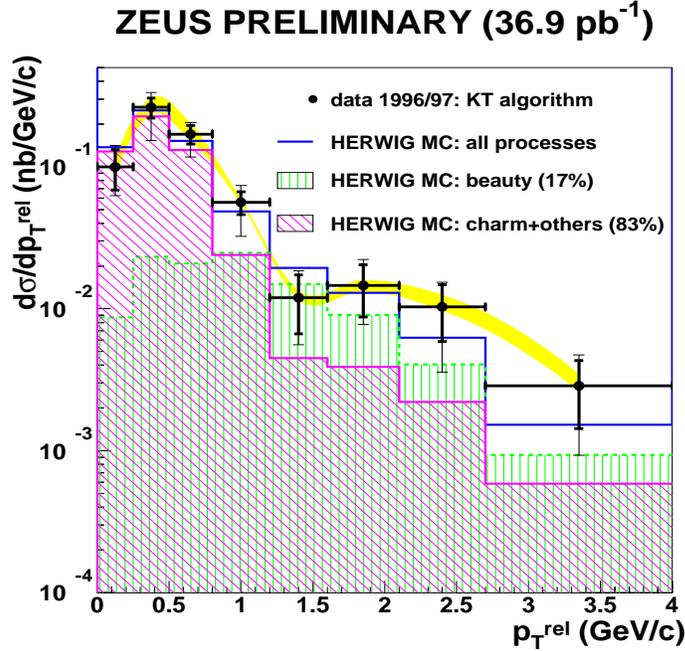}}
\caption[dummy]{\small \label{pt-sl}
The $p_T^{\rm rel}$ distribution for tagged semileptonic decays 
as meaured in ZEUS. The points are ZEUS preliminary
1996 and 1997 data. The inner error bars are the statistical
errors and the outer error bars are the statistical and systematic
errors added in quadrature. The shaded curve represents the
uncertainty due to the energy scale of the jets. The shaded histograms 
show the direct and
resolved contributions for charm, beauty and light quarks 
as predicted by HERWIG 5.9. The solid line shown is the
sum of the contributions.}
\end{figure}

ZEUS have also released a preliminary result of a
beauty measurement by the study of semileptonic decays to electrons. 
The measurement is an inclusive electron
measurement, $e^+p \rightarrow e^- + {\rm dijets} + X$ in the
kinematic region; $Q^2 < 1~{\rm GeV}^2$, $0.2<y<0.8$, two jets with 
$E_T > 7,6$ GeV and $|\eta | < 2.4$ and an electron in the final state 
with $p_T > 1.6$ GeV and $|\eta | < 1.1$. This measurement includes the 
contributions from all beauty, charm and light-quark
decays, but with conversion electrons removed. 
The $x_\gamma^{\rm OBS}$ distribution (Figure~\ref{xg-sl})
shows, similar to the $D^*$ study, a clear tail to low $x_\gamma^{\rm
OBS}$, which can only be described by a significant fraction of 
resolved LO Monte Carlo. The $x_\gamma^{\rm OBS}$ distribution also peaks at
high values, which is consistent with the observation of direct
processes. The preliminary fit on the fraction of the LO resolved
contribution yields $35 \pm 6$\% and agrees well with the
HERWIG~5.9~\cite{herwig} prediction of 40\%. The agreement in shape
between the data and the LO Monte Carlo is good.

A constrained fit was done to the 
$p_T^{\rm rel}$ distribution allowing the fraction of beauty to
vary with respect to the charm and light quarks contribution (fixed in
the ratios given by HERWIG). ZEUS define $p_T^{\rm rel}$ as the momentum of
the electron transverse to the jet axis. A beauty fraction of 
$20\pm 6^{+12}_{- 7}$\% is needed to fit the data (assuming a 40\% resolved
component), in agreement with the HERWIG prediction of $17$\%. The HERWIG 
prediction is about a factor 4 smaller than the data (using 
GRV-LO~\cite{GRVg} for the photon structure function and a beauty mass of
4.95 GeV).
ZEUS quotes a preliminary visible cross section for beauty production
of $\sigma^{\rm vis}_{b\overline{b}}{}(e^+p \rightarrow e^- + {\rm
dijet} + X) = 39\pm 11 ^{+23}_{-16}~\mbox{pb}$ in the kinematic
region described above.

\end{document}